\documentclass[a4paper,11pt]{article}
\pdfoutput=1 

\usepackage{jcappub} 

\usepackage[T1]{fontenc} 

\usepackage{soul,xcolor}
\usepackage{tikz}
\usepackage{xparse}

\newcommand{\highlight}{\color{violet}}

\NewDocumentCommand{\xarrows}{ O{}O{} }{%
\mathrel{%
\vcenter{\hbox{%
\begin{tikzpicture}
  \node[minimum width=1cm,minimum height=1ex,anchor=south,align=center] (a){\text{\vphantom{hg}#1}\\[0.5ex] \vphantom{hg}#2};
  \draw[->] ([yshift=0.35ex]a.west) -- ([yshift=0.35ex]a.east);
  \draw[<-] ([yshift=-0.35ex]a.west) -- ([yshift=-0.35ex]a.east);
\end{tikzpicture}
}}%
}%
}

\def\MNRAS{\emph{Mon. Not. Roy. Astron. Soc. }}

\title{\boldmath Cored density profiles in the DARKexp model}

\author{Claudio Destri}
\affiliation{Dipartimento di Fisica G. Occhialini, Universit\`a
Milano-Bicocca\\ and INFN, sezione di Milano-Bicocca, Piazza della Scienza 3,
20126 Milano, Italia.}
\emailAdd{claudio.destri@mib.infn.it}

\abstract{The DARKexp model represents a novel and promising attempt to solve a long standing
  problem of statistical mechanics, that of explaining from first principles the quasi--stationary
  states at the end of the collisionless gravitational collapse. The model, which yields good fits
  to observation and simulation data on several scales, was originally conceived to provide a
  theoretical basis for the $1/r$ cusp of the Navarro--Frenk--White profile. In this note we show
  that it also allows for cored density profiles that, when viewed in three dimensions, in the
  $r\to0$ limit have the conical shape characteristic of the Burkert profile. It remains to be
  established whether both cusps and cores, or only one of the two types, are allowed beyond the
  asymptotic analysis of this work.}

\begin{document}
\maketitle
\flushbottom

\section{Introduction and summary}\label{sec:intro}
The onset and properties of the quasi--stationary states that emerge from collisionless
gravitational collapses still represent an open problem in cosmology, astrophysics and
statistical mechanics. This concerns dark matter halos, galaxies, galaxy clusters and
stellar systems, spanning several order of magnitudes and different types of
constituents. Most importantly, it concerns the capability itself of statistical physics
to effectively describe the (quasi--)equilibrium of systems with long range interactions
such as Newtonian attraction, which spoils simple additivity and allows true
thermalization through close encounters only on time scales that diverge with the number
of particles \cite{bt}.  In such a situation, the gravitational collapse is well described
\cite{braun,cha3} in the single particle phase--space by the Vlasov--Poisson (or
collisionless Boltzmann) equation, which at the fine--grained level retains full memory of
the initial conditions. The problem is how to reconcile this fact with the observation
that self--gravitating systems of many different kinds and sizes appear, at the
coarse--grained level, to be in very similar quasi--stationary states, as confirmed also
by the remarkable universality features of relaxed structures in $N-$body simulations
\cite{bt,padma,aars,navarroetal}.

Fifty years ago \cite{lyndenb}, Lynden-Bell argued that the coarse--grained phase--space mass
density does rapidly reach a quasi--stationary state through violent relaxation, that is
phase--mixing equilibration driven by the rapid mean--field fluctuations that take place during the
collapse. He also proposed a quantitative description of that quasi--stationary state, obtained
through the maximization of the entropy computed by counting all micro-states compatible with the
Vlasov--Poisson conservation laws. However, this approach is known to have several flaws: it leads
to phase--space densities with infinite mass and energy, even if mass and energy are the first
Vlasov--Poisson constraints; it yields unphysical mass segregation and mass density profiles at odd
with observations, which later turned out to be incompatible also at small scales with the results
of $N-$body simulations \cite{aradjon,aradlyn}.

Many attempts have been made to remedy or at least alleviate the shortcomings of the original
Lynden--Bell's proposal, by invoking incomplete violent relaxation \cite{stiber}, relaxation in a
finite volume \cite{hjomad,chasom}, explicit scattering processes \cite{speher}, change of the
entropy functional to be optimized \cite{plapla}, of the state space over which the
Boltzmann--Shannon entropy is to be computed \cite{hjowill1}, diverse implementations of the
Vlasov--Poisson constraints in maximum entropy calculations \cite{whinar,hekang,pongov} and more.

Here we deal with the DARKexp model, an original and interesting proposal recently put forward by
Hjorth and Williams in \cite{hjowill1}, and further developed and analyzed in
\cite{hjowill2,hjowill3,hjowill4,hjowill5}, with the purpose of providing a statistical--mechanical
basis for the Navarro--Frenk--White density profile \cite{nfw}. The two starting points of the model
are: i) due to the absence of collisions, the distribution function that describes a fully relaxed,
spherically symmetric system, should maximize the standard Boltzmann--Shannon entropy in energy
space rather than in phase space; ii) the small occupation numbers of energy states close to the
bottom of the potential well should be estimated more accurately than with the Stirling
approximation, as done for instance in finite--mass collisional systems near the escape energy 
\cite{madsen}. Eventually, the DARKexp model consists in requiring that the differential energy
distribution \cite{bt} has the form of a lowered Boltzmann exponential
\begin{equation*}
  n(E) \propto \exp[-\beta(E - \Phi(0))] - 1 \;,
\end{equation*}
where $\Phi=\Phi(r)$ is the spherically symmetric gravitational potential, $E=\frac12 v^2+\Phi$ is
the energy of a unit--mass particle and $\beta$ is a Lagrange multiplier that fixes the mean energy
of the system, while the overall normalization is fixed by the total mass. Clearly the DARKexp
$n(E)$ by construction implies a finite mass and a finite energy for the system. Moreover, it is
linear in $E - \Phi(0)$ for $E$ close to $\Phi(0)$, a property that in \cite{hjowill1} is held
responsible for the $1/r$ cusp of the NFW profile. We will examine this crucial point in section
\ref{sec:dark}, where it is shown that such linearity is compatible also with cored profiles.

In fact, the purpose of the present work is to investigate the DARKexp model purely from the
analytical point of view, with a minimum of assumptions and without resorting to numerical
approximations. In particular, we concentrate our attention on the neighborhood of $E = \Phi(0)$,
that is to the center of the system, where we perform an accurate asymptotic analysis. In this
respect, our findings do not hinder the fitting of DARKexp predictions to observations and $N-$body
simulations, which is reported to be quite good
\cite{hjowill3,beraldo13,umetzu16,nolting16}. Indeed, no cusp/core ambiguity may arise when the
DARKexp $n(E)$ is directly compared to the differential energy distribution extracted from $N-$body
simulations \cite{hjowill3,nolting16}, while the direct comparison of the DARKexp mass density,
numerically computed as in \cite{hjowill2}, to that of astrophysical objects
\cite{beraldo13,umetzu16} and of $N-$body simulations \cite{nolting16} are restricted to distances
too large to distinguish a cusp from a small core.

\smallskip 
In the context of cosmic small--scale structure, the dark matter (DM) cusp--core problem has a long
and strongly debated history. While $N-$body simulations support cuspy profiles for pure cold DM
halos, albeit not exactly of NFW type \cite{navarro2004,ludlow2013}, direct observations, especially
of DM--dominated galaxies, seem to favor cored DM halos \cite{salucci2007,walker2011,adams2014}.
Cusps might also be turned into cores by baryonic feed--backs
\cite{navarro1996,governato2012,pontzen2012,chan2015,brooks2017}, whilst warm DM does support cores
by itself, although apparently smaller than the observed ones
\cite{villaes2011,maccio2013,lovell2014}. Also elastic collisions of self--interacting CDM can alter
the collisionless density profiles, improving agreement with observations over a wide range of
galaxy masses \cite{spergel2000,vogelsberger2012,rocha2013,zavala2013,vogelsberger2016}. In this
controversy, the DARKexp model has been regarded as a first--principle assist to cusps, although DM
halo formation, being certainly constrained by the cosmological context, cannot rely purely on
equilibrium statistical mechanics. In this note we show that the DARKexp model is, just by its
definition, at most neutral in the cusp vs. core contest. If we must express a preference, it is
probably for cores, since our asymptotic analysis exhibits troublesome features in the case
of $1/r-$cusps, which are totally absent in the case of cores.

\medskip This work is organized as follows. In the next section (sec. \ref{sec:dless}) we set up a
dimensionless framework for the description of $1/r-$cusped and cored profiles in the context of
ergodic systems, that is systems whose phase--space mass density $f$ depends on position and
velocity only through the unit--mass one--particle energy $E=\frac12 v^2+\Phi$. In section
\ref{sec:dark} we formulate in very concise dimensionless terms the main problem of the DARKexp
model, that is the determination of a distribution $f(E)$ that produces the given differential
energy distribution $n(E)$. Section \ref{sec:corc} is devoted to a criticism of the argument in
\cite{hjowill1}, section 3.1, which restricts to $1/r-$cusped profiles the possibility that $n(E)$
is linear in $E - \Phi(0)$ as $E\to\Phi(0)$. In section \ref{sec:sublead} we present, both for cusps
and cores, a recursive asymptotic analysis to obtain the subleading corrections to $f(E)$ as
$E\to\Phi(0)$; in the case of cores it is possible to determine the analytic structure and the first
two analytic corrections up to a normalization [see eqs.~\eqref{eq:asympt} and~\eqref{eq:asycore},
where $s\propto E-\Phi(0)$], while in the case of $1/r$ cusps one must face an infinite
proliferation of logarithmic divergent terms; we then verify that these logarithms cannot resum into
a simple non--integral power; rather, if they do resum, it must be into a non--analytic term
oscillating in $\log(E - \Phi(0))$. Finally, in section \ref{sec:corig} we derive the analytic
structure of the cored density near the origin. It can be written as
\begin{equation*}
  \rho(r) \simeq \rho(0)\left[1 + \gamma_0r + \gamma_1r^2\log r + O(r^2)\right]\; ,
\end{equation*}
where
\begin{equation*}
  \gamma_0=\frac{\rho'(0)}{\rho(0)},\quad  \gamma_1 = - \frac{\!\!32}{5\pi^2}\,\gamma_0^2
\end{equation*} 
and the two dimension--full quantities $\rho(0)>0$ and $\gamma_0<0$ determine the two free scales of
the system. This profile is cored because the logarithmic slope $r\rho'(r)/\rho(r)$ vanishes
linearly at the origin. Quite interestingly, when viewed in three dimensions, the mass density
exhibits a conical shape in the neighborhood of the origin as the Burkert profile \cite{burkert95},
extensively used to yield good fits to rotation curves in both dwarfs and spiral galaxies (see for
instance \cite{salucci2012}).

It should be stressed that, from a rigorous mathematical point of view, our findings cannot be
regarded as proving or disproving the existence of the DARKexp $f(E)$, either for cored or for
$1/r-$cusped density profiles. They are just statements on the asymptotics of the solution $f(E)$ as
$E\to\Phi(0)$, \emph{assuming} that such a solution exists. Strictly speaking, it is not obvious
that a solution exists for any given value of $\beta$ or that it is unique when it exists. At the
moment, the only way to find a solution seems to be through some numerical method that, unlike the
simple iteration scheme of refs. \cite{binney82,hjowill1}, does keeps fully into account the leading
asymptotic behavior of $f(E)$ as $E\to\Phi(0)$, whether it corresponds to cores or cusps. We plan to
report on such a numerical approach in a forthcoming publication.

\section{Dimensionless formulation} \label{sec:dless} 

Let $\rho(r)$ be the spherically symmetric mass density. The gravitational potential
$\Phi(r)$ then solves Poisson's equation
\begin{equation}\label{eq:pois0}
\frac1{r^2} (r^2\Phi')' = \Phi''+\frac2r \Phi' = 4\pi G\rho
\end{equation}
which implies Gauss's law
\begin{equation*}
  \Phi'(r) = G\frac{{\cal M}(r)}{r^2}\;,\quad {\cal M}(r)=4\pi\!\int_0^r\!dr\,r'^2\rho(r')
\end{equation*}
and therefore $\Phi'(r)>0$. We assume that the total mass $M=\lim_{r\to\infty}{\cal M}(r)$ is
finite. Then we can also assume $\Phi(r)<0$ everywhere, with $\Phi(0)$ finite and $\Phi(r)\to 0$ as
$r\to\infty$. This is the standard physically sensible framework.

If, besides a finite mass, the system has also a finite size, then there exist a $r_0>0$ such that
$\rho(r)=0$ for $r\ge r_0$. Of course, we can include also the case of infinite size by letting
$r_0\to\infty$. Clearly, the shifted potential $\phi=\Phi+G M/r_0$ still satisfies Poisson's
equation and vanishes at $r=r_0$.

We consider two physically relevant cases for $\rho(r)$ and $\phi(r)$:
\begin{description}
\item{(A) cored profiles:} \\
  $\rho_0\equiv\rho(0)$ is finite, $\phi(r)\simeq \phi(0) +\frac23\pi G \rho_0\,r^2$ as $r\to 0$;
\item{(B) profiles with a $1/r$ cusp in the mass density:}\\
  $\phi(r)\simeq \phi(0) + \phi'(0)\,r$ as $r\to 0$, with $\phi'(0)>0$; then $\rho(r)\simeq
  \frac{\phi'(0)}{2\pi G} r^{-1}$ as $r\to 0$.
\end{description}
In both cases we may set
\begin{equation*}
   \rho(r) = \frac{|\phi(0)|}{4\pi G\,r_\ast^2}\, \nu(u(x))\;,\quad \phi(r)=\phi(0)[1-u(x)]
   \;,\quad x\equiv\frac{r}{r_\ast}\;,
\end{equation*}
where $\nu(u)$ and $u(x)$ are dimensionless function of their own dimensionless argument and
\begin{equation*}
   {\rm (A):} \quad r_\ast^2=\frac{3|\phi(0)|}{4\pi G\rho_0}\quad ;\qquad
   {\rm (B):} \quad r_\ast=\frac{|\phi(0)|}{\phi'(0)}\;.
\end{equation*}
Notice that, by construction, as $x$ grows from $0$ to $x_1=r_0/r_\ast$, $u(x)$ monotonically grows
from $0$ to $1$. To complete this dimensionless setup we assume units such that $4\pi G = 1$.

Poisson's equation now reads
\begin{equation}\label{eq:pois1}
   u''(x) +\frac2x u'(x) = \nu(u(x))
\end{equation}
and we look for solutions which behaves as follows near $x=0$:
\begin{equation}\label{eq:asy}
   {\rm (A):} \quad u \simeq \frac12 x^2 \;,\quad \nu \simeq 3\quad ;  \qquad 
   {\rm (B):} \quad u \simeq x\;, \quad \nu \simeq \frac2x\;.
\end{equation}
In both cases we also have
\begin{equation}\label{eq:phiphi}
  M = 4\pi r_\ast|\phi(0)|\,m \;,\quad \Phi(0) = \Bigl(1+\frac{m}{x_1}\Bigr)\phi(0) \;, 
\end{equation}
where
\begin{equation*}
  m = \int_0^{x_1}dx\, x^2\nu(u(x)) \;,\quad x_1=\frac{r_0}{r_\ast} \;.
\end{equation*}
are the dimensionless mass and size of the system, respectively. Changing integration variable from
to $x$ to $u$ we obtain
\begin{equation*}
  m = \int_0^1du\, x^2(u)x'(u)\nu(u) \;.
\end{equation*}
where the function $x(u)$, the inverse of $u(x)$, fulfills the transformed Poisson's equation
\begin{equation}\label{eq:pois2}
   \frac{2}{x(u) x'(u)} - \frac{x''(u)}{[x'(u)]^3} = \nu(u) \;,
\end{equation}
which is to be solved with the behavior implied by eq.~\eqref{eq:asy} as $u\to0$, that is 
\begin{equation}\label{eq:asy2}
   {\rm (A):} \quad x \simeq (2u)^{1/2} \;,\quad \nu \simeq 3\quad ;  \qquad 
   {\rm (B):} \quad x \simeq u\;, \quad \nu \simeq \frac2u\;.
\end{equation}
Two (small) advantages of eq.~\eqref{eq:pois2} over the more common eq.~\eqref{eq:pois1} are: 
\begin{description}
\item{(i)} the non-linearity is fixed independently on the form of $\nu(u)$; 
\item{(ii)} the domain of the independent variable is fixed once and for all to $[0,1]$, while the
  system is bounded iff $x_1=\lim_{u\to1}x(u)$ is finite.
\end{description}
Moreover, with $u$ as independent variable, we can write Gauss's law in the dimensionless form
\begin{equation*}
  \int_0^u\!ds\,x^2(s)x'(s)\nu(s) = \frac{x^2(u)}{x'(u)}
\end{equation*}

\medskip
Now let $f(\frac12 v^2+\phi(r))$ be the ergodic phase--space mass distribution that produces the
density $\rho(r)$, that is
\begin{equation}\label{eq:f2rho}
  \begin{split}
    \rho(r) &= \int\!d^3v\,f(\tfrac12 v^2+\phi(r)) = \int\!dE\,f(E)\int d^3v\, 
    \delta(\tfrac12 v^2+\phi(r)-E)\\
    &= 4\pi \int_{\phi(r)}^0 \!dE\,f(E)\,[2(E-\phi(r))]^{1/2}\;,
  \end{split}
\end{equation}
where we have used that $f(E)=0$ for $E>0$ as dictated by the condition of bound matter with finite
total mass. Notice that we have taken the unit--mass one--particle energy to be $E=\frac12 v^2+\phi$
rather than $\frac12v^2+\Phi$. Hence $E$ assumes values between $\phi(0)$ and $0$ rather than
$\Phi(0)$ and $-GM/r_0$, regardless of the size of the system.

Furthermore, let $n(E)$ be the mass density of states with a given
one--particle energy $E$, or differential energy distribution \cite{bt}, namely
\begin{equation*}
   n(E) = \frac{dM}{dE} = \int\!d^3r\!\int\!d^3v\,f(\tfrac12 v^2+\phi(r)) 
   \delta(\tfrac12 v^2+\phi(r)-E) = f(E)g(E) \;,
\end{equation*}
where $g(E)$ is the kinematic density of states
\begin{equation}\label{eq:g}
    g(E) = \int\!d^3r\!\int\!d^3v\,\,\delta(\tfrac12 v^2+\phi(r)-E) =
    (4\pi)^2 \int_0^{r_E} \!dr\,r^2\,[2(E-\phi(r))]^{1/2}\;.
\end{equation}
and $r_E$ is the radius where the one--particle kinetic energy vanishes, that is $E=\phi(r_E)$.  By
definition
\begin{equation}\label{eq:M}
  M = \int_{\phi(0)}^0\!dE\,n(E) \;.
\end{equation}
If we parametrize the one--particle energy as the potential, that is 
\begin{equation}\label{eq:s2E}
  E=\phi(0)(1-s) \;,\quad 0\le s\le 1 \;,
\end{equation}
we can express $f(E)$ and $g(E)$ in terms of two dimensionless function $F(s)$ and $G(s)$
as
\begin{equation*}
    f(E) = \frac{(2|\phi(0)|)^{-1/2}}{4\pi\,r_\ast^2}\,F(s) \quad ;\qquad
    g(E) = (4\pi)^2\,r_\ast^3\,(2|\phi(0)|)^{1/2} \,G(s) \;.
\end{equation*}
Then
\begin{equation*}
   n(E) = 4\pi\,r_\ast\, N(s) \;,\quad  m = \int_0^1\!ds\,N(s)\;,\quad N(s) = F(s)G(s)
\end{equation*}
while eq.~\eqref{eq:f2rho} implies
\begin{equation}\label{eq:F2nu}
   \nu(u) = \int_u^1\!ds\, F(s)\,(s-u)^{1/2} \;.
\end{equation}
Changing integration variable from $r$ to $x$ and then to $u$ in eq.~\eqref{eq:g}, we obtain also
\begin{equation}\label{eq:x2G}
   G(s) = \int_0^s\!du\,x^2(u)x'(u)\,(s-u)^{1/2} = \frac16 \int_0^s\!du\,x^3(u)\,(s-u)^{-1/2}\;.
\end{equation} 
According to one of Abel's theorems, the half--primitives in eqs.~\eqref{eq:F2nu} and \eqref{eq:x2G}
can be inverted in terms of half--derivatives
\begin{equation}\label{eq:nu2F}
  F(s) = \frac{2}{\pi}\frac{d}{ds}\int_s^1\!du\, \nu'\!(u)\,(u-s)^{-1/2}
\end{equation}
and
\begin{equation}\label{eq:G2x}
  x^3(u) = \frac{6}{\pi}\int_0^u\!ds\, G'(s)\,(u-s)^{-1/2} \;.
\end{equation}
The various relations between the functions $F$, $\nu$, $x$ and $G$ can be summarized as
\begin{equation}\label{eq:chain}
  F(s)\, \xarrows[\small eq.~\eqref{eq:F2nu}][\small eq.~\eqref{eq:nu2F}]\,\nu(u)
  \,\xarrows[\small eq.~\eqref{eq:pois2}][\small eq.~\eqref{eq:pois2}] \,x(u)\,
    \xarrows[\small eq.~\eqref{eq:x2G}][\small eq.~\eqref{eq:G2x}] \,G(s) \;.
\end{equation}
Clearly, $F$ and $G$ are complicated non--local and non--linear functionals one of the other.

\section{The DARKexp model}\label{sec:dark}

As anticipated in the Introduction, the DARKexp model \cite{hjowill1} consists in assuming that the
differential mass--to--energy distribution is proportional to $\exp[-\beta(E - \Phi(0))] - 1$ if 
$E=\frac12 v^2+\Phi$ is the unit--mass one--particle energy. With our convenience choice of 
$E=\frac12 v^2+\phi$, the DARKexp differential distribution has the form
\begin{equation*}
  n(E) = C\, \{\exp[-\beta(E - \phi(0))] - 1\} \quad,\qquad \phi(0) \le E\le 0 \;.
\end{equation*}
The two dimension--full constants $C$ and $\beta$ act as Lagrange multipliers that fix the total
mass and the average one--particle energy, respectively. In the dimensionless setup, we set
\begin{equation*}
  b = \beta \phi(0) \;,\qquad C' = \frac{C}{4\pi\,r_\ast} 
\end{equation*}
so that 
\begin{equation}\label{eq:FG}
  N(s) = F(s)G(s) = C'\, (e^{bs} - 1) = \frac{mb}{e^b-1-b}\,(e^{bs} - 1) \;,
\end{equation}
where the last equality, fixing $C'$, ensures that $\int_0^1\!ds\,N(s)=m$. $N(s)$ is
positive--defined, as necessary, both for $b<0$ and $b>0$. In both cases it is also monotonically
increasing in $s$, that is in the one--particle energy $E$. On the other hand, $N(s)$ is convex for
$b>0$ and concave for $b<0$, while it reduces to a linear ramp for $b=0$. Clearly, when $b\ge0$ or
$b<0$ the ``inverse temperature'' $\beta$ is respectively negative, null, or positive. However, the
terminology ``inverse temperature'' is rather improper: in this context $\beta$ is just a Lagrange
multiplier.

In order to relate the parameter $b$ to the parameter $\phi_0=\beta\Phi(0)$ of
ref.~\cite{hjowill1}, one must keep in mind the shift by $GM/r_0$ performed here on the
one--particle unit--mass energy and the gravitational potential
w.r.t. ref.~\cite{hjowill1}. Using the second of eqs.~\eqref{eq:phiphi}, one readily gets
\begin{equation*}
  \phi_0 =  \Bigl(1+\frac{m}{x_1}\Bigr)b
\end{equation*}
which actually gives $\phi_0$ in terms of $b$ only, since both $m$ and $x_1$ are functions of $b$ only, 
as will become clear later on.

The dimensionless parameter $b$ fixes the average value of the one--particle energy,
$\bar{E}=\phi(0)(1-\bar{s})$, where
\begin{equation*}
  \bar{s} = \frac{\int_0^1\!ds\,s(e^{bs}-1)}{\int_0^1\!ds\,(e^{bs}-1)} = 
  \frac{(b-1)e^b+1-b^2/2}{b(e^b-1-b)}\;.
\end{equation*}
As $b$ ranges from $-\infty$ to $+\infty$, $\bar s$ monotonically grows from $1/2$ to $1$, so that a
solution for $b$ exists only for $\bar{s}>1/2$ and is unique. Moreover, $b(\bar{s})$ is negative for
$1/2<\bar{s}<2/3$ and positive for $\bar{s}>2/3$.%

\medskip
The problem is now to determine the explicit form (or forms) of $F(s)$ and $G(s)$, taking into
account the rather involved functional relation, eq.~\eqref{eq:chain}, that connects the two.

\medskip
Eq.~\eqref{eq:FG} implies that $F(s)G(s) \sim s$ as $s\to0$. Recalling eq.~\eqref{eq:asy}, we also
have for cores or cusps, respectively, as $u\to0$ and $s\to0$
\begin{equation}\label{eq:asy3}
  \begin{split}
    &{\rm (A):} \quad x^2x' \simeq (2u)^{1/2} \quad\Longrightarrow\quad 
    G(s) \simeq \tfrac{1}{4\sqrt2}\pi s^2 \;; \\[2ex]
    &{\rm (B):} \quad x^2x' \simeq u^2 \quad\Longrightarrow\quad 
    G(s) \simeq \tfrac{16}{105}s^{7/2} \;.
 \end{split}
\end{equation}
where we used the definition of Euler's Beta function
\begin{equation*}
  \int_0^s\!du\,u^{z-1}(s-u)^{w-1} = B(z,w)s^{z+w-1} 
\end{equation*}
along with $B(\tfrac32,\tfrac32)=\tfrac18\pi$ and $B(3,\tfrac32)=\tfrac{16}{105}$. Hence
\begin{equation}\label{eq:asy4}
   {\rm (A):} \quad F(s) \simeq K s^{-1} \quad;\qquad  
   {\rm (B):} \quad F(s) \simeq K s^{-5/2}\;.
\end{equation}
for some constant $K$ to be determined. For cores, when $u=0$, the $s^{-1/2}$ singularity in the
integrand of eq.~\eqref{eq:F2nu} is integrable, while the $s^{-5/2}$ behavior of $F(s)$, in the case
of cusps, yields the $u^{-1}$ behavior of $\nu(u)$. Matching with the asymptotic condition for
$\nu(u)$ in eq.~\eqref{eq:asy2} then fixes $K$ only for the (B) case of cusps and in particular
$K=3$, since
\begin{equation*}
  \int_u^1\!ds\,s^{-5/2}(s-u)^{1/2} = \tfrac23 u^{-1}(1-u)^{3/2} \simeq \tfrac23 u^{-1}\;,\quad u\to0\;.
\end{equation*}
In the (A) case of cores, instead, all values of $F(s)$ over the range $0\le s\le 1$ contribute to
$\nu(0)=3$ and $K$ cannot be determined without solving the full problem.

\section{The flawed argument against cores}\label{sec:corc}

In refs.~\cite{hjowill1} only case (B), that is $1/r-$cusps, was considered. The (flawed)
argument used in ~\cite{hjowill1} to exclude any other possibility would proceed as
follows in the present dimensionless formulation.

Suppose $u\sim x^\alpha$ as $x\to0$, then Poisson's eq.~\eqref{eq:pois1} implies 
\begin{equation*}
  \nu\sim x^{\alpha-2} \sim u^{1-2/\alpha} \;.
\end{equation*}
Now suppose
\begin{equation*}
  F(s)\sim s^{-1/2-2/\alpha}
\end{equation*}
as $s\to0$. Then, by eq.~\eqref{eq:F2nu}, we have 
\begin{equation*}
   \nu(u) \sim \int_u^1\!ds\, s^{-1/2-2/\alpha}\,(s-u)^{1/2} = u^{1-2/\alpha} B(1-u;\tfrac32,1-2/\alpha)
\end{equation*}
in terms of the incomplete Beta function
\begin{equation*}
  B(y,z,w) = \int_0^y\!dt\,t^{z-1}(1-t)^{w-1}\;. 
\end{equation*}
As $u\to0$, one recovers the complete Beta $B(\tfrac32,1-2/\alpha)$ which exists finite
as long as $2/\alpha$ is not a natural number. Now from eq.~\eqref{eq:x2G} we obtain
\begin{equation*}
  G(s) \sim  \int_0^s\!du\,u^{3/\alpha}\,(s-u)^{-1/2} \sim s^{1/2+3/\alpha}
\end{equation*}
and therefore
\begin{equation*}
  F(s)G(s) \sim s^{1/\alpha} \;,
\end{equation*}
which is compatible with the DARKexp $N(s)$ only for $\alpha=1$, that is for case (B) of
$1/r-$cusps.
 
This conclusion is indeed correct as long as $\alpha<2$, but does not necessarily hold for
$\alpha=2$, because of the logarithmic divergence at $u=0$ of the incomplete Beta
$B(1-u,\tfrac32,0)$. Indeed, if we take $\alpha=2$ to mean that $\nu(0)$ is finite, rather than
$\nu(u)$ logarithmically divergent as $u\to 0$, then what must vanish as $\alpha\to2$ is the
normalization constant in $F(s)$ [the extension to generic $\alpha$ from $\alpha=1$ or $\alpha=2$ of
the constant $K$ of eq.~\eqref{eq:asy4}] that enforces the finite value of $\nu(0)$, say
$\nu(0)=3$. Then $F(s)\sim s^{-1/2-2/\alpha}$ will not hold at $\alpha=2$. Since $G(s)\sim s^2$ for
any cored density profile, we see that $F(s)\sim s^{-1}$ provides a solution to $F(s)G(s)\sim s$,
which is exactly the one corresponding to case (A) above.

\smallskip The central question then becomes whether both case (A) and (B) are allowed, or only one
of them, when we go beyond the leading term in the asymptotic behavior as $s,u\to0$, for instance by
considering the whole unit interval for $s$ and $u$. This complete solution requires a numerical
analysis which is beyond the scope of this work will be the subject of another report. Here we focus
our attention on the subleading terms in the asymptotic expansions near $s=0$ and $u=0$. In other
words, we try to go beyond the leading terms:
\begin{equation}\label{eq:lead} 
  \begin{split}
    &{\rm (A):} \quad f(s) \simeq Ks^{-1}\;,\quad \nu(u) \simeq 3 \;, \quad
    x(u) \simeq (2u)^{1/2}\;, \quad G(s) \simeq \tfrac1{4\sqrt2}\pi s^2 \,; \\[2ex]
    &{\rm (B):} \quad f(s) \simeq 3\,s^{-5/2}\;,\; \nu(u)\simeq 2\,u^{-1} \;,\quad
    x(u)\simeq u \;,\quad    G(s) \simeq \tfrac{16}{105}s^{7/2} \;.
  \end{split}
\end{equation}

\section{A convenient  reformulation}

Let us reformulate the DARKexp problem in a more convenient way. We set 
\begin{equation}\label{eq:hhhh} 
  \begin{split}
    &{\rm (A):} \quad 
    \begin{cases}
      \;F(s) = Ks^{-1}[1+h_F(s)] \;,\; & \nu(u) = 3[1+h_\nu(u)]  \\[1ex]
      \;x(u) = (2u)^{1/2}[1+h_x(u)]  \;,\; & G(s) = \tfrac{\pi}{4\sqrt2}s^2[1+h_G(s)] \;;
    \end{cases}
    \\[2ex] 
    &{\rm (B):} \quad
    \begin{cases}
      \;F(s) = Ks^{-5/2}[1+h_F(s)] \;,\; & \nu(u) = 2u^{-1}[1+h_\nu(u)]  \\[1ex]
      \;x(u) = u[1+h_x(u)] \;,\; &  G(s)=\tfrac{16}{105}s^{7/2}[1+h_G(s)]\;;
    \end{cases}
  \end{split}
\end{equation}
In both cases, by eqs.~\eqref{eq:FG} the functions $h_F(s)$ and $h_G(s)$, which are related by the
functional chain in eq.~\eqref{eq:chain}) through $h_\nu(u)$ and $h_x(u)$, must satisfy
\begin{equation}\label{eq:hhh2}
   [1+h_F(s)][1+h_G(s)] = \frac{e^{bs}-1}{bs}
\end{equation}
while
\begin{equation}\label{eq:Km}
  \begin{split}
    &{\rm (A):} \quad K = \frac3{\int_0^1\!ds[1+h_F(s)]s^{-1/2}}\;,
    \quad m = \frac{\pi}{4\sqrt2}\frac{e^b-1-b}{b^2}K\;;\\[2ex]
    &{\rm (B):} \quad K = 3\;,\quad m = \frac{16}{35}\frac{e^b-1-b}{b^2} \;.
  \end{split}
\end{equation}
The leading asymptotics \eqref{eq:lead} is now rewritten as
\begin{equation}\label{eq:hhh3}
  \lim_{s\to0}h(s)=0  \;, \quad h = h_F,\,h_\nu,\,h_x,\,h_G \;.
\end{equation}
This last condition could be relaxed by accepting that the $h(s)$ functions could
indefinitely oscillate around $0$ as $s\to0$, still remaining bounded and preserving
positivity, that is $1+h(s)>0$. An example of such a behavior is provided by
$a\sin(\log(s))$ with $|a|<1$.  This would mean that the leading asymptotic behavior is
true only ``on average''. In the following section we conservatively assume eq.~\eqref{eq:hhh3}
and try to obtain asymptotic expansions of the $h(s)$ functions in both cases (A) and (B).

\section{Subleading asymptotics}\label{sec:sublead}

We consider the chain \eqref{eq:chain}) from left to right (one could proceed in other ways with
the same conclusions) and first check what subleading terms in $h_\nu(u)$, $h_x(u)$ and $h_G(s)$ are
entailed by the leading form of $F(s)$, namely $h_F(s)\simeq 0$. Then we use the first order
expansion of eq.~\eqref{eq:hhh2}, that is
\begin{equation}\label{eq:ntl} 
  h_F(s) \simeq \tfrac12 bs -  h_G(s) \;,
\end{equation}
to compute the correction to $h_F(s)$ and then restart the process. Since the first step
\eqref{eq:F2nu} of the chain plays a crucial role, we rewrite it here explicitly for $h_F(u)$ and
$h_\nu(u)$:
\begin{equation}\label{eq:hF2hnu} 
  \begin{split}
    &{\rm (A):} \quad h_\nu(u) = \tfrac13 K\left\{\int_u^1\!ds\, [1+h_F(s)]\,s^{-1}(s-u)^{1/2} 
      - \int_u^1\!ds\, [1+h_F(s)]\,s^{-1/2} \right\}
      \;; \\[2ex]
    &{\rm (B):} \quad h_\nu(u) = \tfrac32 u\!\int_u^1\!ds\, [1+h_F(s)]\,s^{-5/2}\,(s-u)^{1/2}-1 \;.
  \end{split}
\end{equation}

\smallskip\noindent
In this analysis we shall repeatedly use the following result
\begin{equation}\label{eq:myint}
  \begin{split}
    \int_u^1\!ds\, s^{z-3/2}\,(s-u)^{1/2} &= \left[-\int_{1/u}^\infty\!dt\,t^{z-3/2}(t-1)^{1/2} +
      \int_1^\infty\!dt\,t^{z-3/2}(t-1)^{1/2} \right]u^z \\ &= z^{-1}
    F\big(-z,-\tfrac12;1-z;u\big) + B(-z,\tfrac32)\,u^z\;,
  \end{split}
\end{equation}
where $z$ is any complex number and the special case of the hypergeometric series
\begin{equation*}
  F\big(-z,-\tfrac12;1-z;u\big) = 
  \sum_{n=0}^\infty (-1)^n\frac{z}{z-n} {{\tfrac12}\choose{n}} u^n
\end{equation*}
uniformely converges in the unit $u$ disk. Notice that the integrals over $t$ converge at
$t\to\infty$ for $\Re z<\tfrac12$, but the result can be analytically continued everywhere else. Notice
also that the pole that appear in the Beta function $B(-z,\tfrac32)$, when $z$ is a non--negative
integer $k$, is canceled by a similar pole in one of the coefficients of the hypergeometric series,
resulting into an overall  $u^k\log u$ term.

The type of integrals needed in eq.~\eqref{eq:x2G}, to compute $h_G(s)$ from $h_x(u)$, are simpler,
since they reduce to direct instances of the Beta function as in section \ref{sec:corc}:
\begin{equation*}
  \int_0^s\!du\, u^{z-1}\,(s-u)^{1/2} = B(z,\tfrac32) \,s^{z+1/2} \;.
\end{equation*}

\smallskip
The first step gives
\begin{equation*}
  \begin{split}
    &{\rm (A):} \quad h_\nu(u) \simeq \tfrac13 K\left[\int_u^1\!ds\, s^{-1}\,(s-u)^{1/2} -
    \int_0^1\!ds\, s^{-1/2}\right] \simeq -\tfrac13\pi K u^{1/2} \;; \\[2ex]
    &{\rm (B):} \quad h_\nu(u) \simeq \tfrac32 u\int_u^1\!ds\, s^{-5/2}\,(s-u)^{1/2} -1 =  
    (1-u)^{3/2} -1 \simeq -\tfrac32 u \;.
  \end{split}  
\end{equation*}
In turn, solving eq.~\eqref{eq:pois2} to the same order yields
\begin{equation*}
    {\rm (A):} \quad h_x(u) \simeq \tfrac{1}{12}\pi K u^{1/2} \quad; \qquad
    {\rm (B):} \quad h_x(u) \simeq \tfrac38 u \;,
\end{equation*}
which substituted in eq.~\eqref{eq:x2G} finally entails
\begin{equation*}
    {\rm (A):} \quad h_G(s) = \tfrac{32}{45} K s^{1/2}\quad; \qquad
    {\rm (B):} \quad h_G(s) = s \;.
\end{equation*}
Then, by eq.~\eqref{eq:ntl}, to leading order we have
\begin{equation*}
  {\rm (A):} \quad  h_F(s)\simeq - \tfrac{32}{45} K s^{1/2}\quad,\qquad
  {\rm (B):} \quad  h_F(s)\simeq \big(\tfrac12 b - 1\big)s \;.
\end{equation*}
Restarting the chain, one obtains
\begin{equation}\label{eq:o2nu}
  \begin{split}
    &{\rm (A):} \quad h_\nu(u) \simeq - \tfrac13\pi K u^{1/2} - \tfrac{16}{135}K^2\,u\log u\;; \\[2ex]
    &{\rm (B):} \quad h_\nu(u) \simeq - \tfrac32 u + \tfrac32\big(\tfrac12 b - 1\big)u\log u\;.
  \end{split}  
\end{equation}
Then
\begin{equation*}
  \begin{split}
    &{\rm (A):} \quad h_x(u) \simeq \tfrac{1}{12}\pi K u^{1/2} + \tfrac4{225}K^2\,u\log u \;;
    \\[2ex]
    &{\rm (B):} \quad h_x(u) \simeq -\tfrac38\big(\tfrac12 b - 1\big)u\log u + 
    \tfrac38 \big[1 + \tfrac32\big(\tfrac12 b - 1\big)\big] u \;,
  \end{split}  
\end{equation*}
and finally 
\begin{equation}\label{eq:o2G}
  \begin{split}
    &{\rm (A):} \quad h_G(s) \simeq  \tfrac{32}{45} K s^{1/2} +\tfrac2{45}K^2\,s\log s \;;
    \\[2ex]
    &{\rm (B):} \quad h_G(s) \simeq - \big(\tfrac12 b - 1\big)s\log s  + 
    \big[1 + \big(\tfrac{5029}{1260}-2\log 2\big)\big(\tfrac12 b - 1\big)\big]\,s \;.
  \end{split}  
\end{equation}
Evidently, this iterative procedure has a serious problem in case (B), since $h_G(s)$ now contains a
$s\log s$ correction which is of {\em smaller} order w.r.t. the first order correction in $h_F(s)$
which generated it. Moreover, the $s$ term in $h_G(s)$ changed w.r.t. the previous iteration,
requiring to change also the $s-$term in $h_F(s)$. One could introduce two unknown coefficients for
the $s\log s$ and $s$ corrections in $h_F(s)$ and use eq.~\eqref{eq:ntl} to try and fix them
together at the end of the chain. This cannot work, however, because the integration in
eq.~\eqref{eq:myint} shows that a $s\log s$ term in $h_F(s)$ would produce a $u\log^2\!u$ term in
$h_\nu(u)$, which would in turn imply a similar term in $h_x(u)$, resulting into a $s\log^2\!s$ in
$h_G(s)$ that would require to add a $s\log^2\!s$ to $h_F(s)$ by \eqref{eq:ntl}. But then a
$s\log^3\!s$ would appear in $h_G(s)$ and so on and so forth. In general, to a $s\log^n\!s$ term in
$h_F(s)$ there corresponds a $s\log^{n+1}\!s$ in $h_G(s)$, making it impossible to satisfy
eq.~\eqref{eq:hhh2} at any finite order in powers of $\log s$.

\smallskip 
Before analyzing in more detail this tricky situation of case (B), let us consider the situation
of case (A). Eq.~\eqref{eq:ntl} now requires to leading order 
\begin{equation*}
  {\rm (A):} \quad  h_F(s)\simeq - \tfrac{32}{45} K s^{1/2} -\tfrac2{45}K^2\,s\log s
\end{equation*}
and the effects of the new $s\log s$ correction must be computed.  But by direct
calculation or by eq.~\eqref{eq:myint} [noticing the zero in $B(-z-\tfrac12,\tfrac32)$
when $z=1$] we now find
\begin{equation}
  \int_u^1\!ds\, (\log s)(s-u)^{1/2} \simeq -\tfrac49 + \tfrac83 u
\end{equation} 
so that both $u^{1/2}$ and $u\log u$ corrections remain valid for $h_\nu(u)$ in
eq.~\eqref{eq:o2nu}. Moreover, at this order the non--linearity of the step \eqref{eq:pois2} does
not mix $u^{1/2}$ with $u\log u$ so that eventually both $s^{1/2}$ and $s\log s$ corrections remain
valid for $h_G(s)$ in eq.~\eqref{eq:o2G}. To take into account the higher order $\tfrac12 bs$ term
in eq.~\eqref{eq:ntl} one could try to add an explicit correction of order $s$ in $h_F(s)$ and
$h_G(s)$:
\begin{equation*}  
  {\rm (A):} \quad  
  \begin{cases}
    h_F(s)\simeq - \tfrac{32}{45} K s^{1/2} -\tfrac2{45}K^2\,s\log s 
    + \big(\tfrac12 b - \lambda \big)  s\;, \\[1ex]
    h_G(s)\simeq \tfrac{32}{45} K s^{1/2} +\tfrac2{45}K^2\,s\log s + \lambda s \;.
  \end{cases}
\end{equation*}
where $\lambda$ could be fixed by the chain computed up to order $s$.  Unfortunately, this procedure
is not stable upon higher order calculations, that is when eq.~\eqref{eq:hhh2} is considered up to
order $s^n$ with $n>1$
\begin{equation}\label{eq:nth} 
  h_F(s)+h_G(s)+h_F(s)h_G(s) \simeq \tfrac12 bs + \tfrac16(bs)^2 + \ldots + \tfrac1{(n+1)!}(bs)^n \;.
\end{equation}
This is because all higher order terms in $h_F(s)$ affect the correction of order $u$ in $h_\nu(u)$
and because the non--linearity of the step \eqref{eq:pois2} and in eq.~\eqref{eq:nth} introduces in
the game all mixed powers of $s^{1/2}$ and $s\log s$. To obtain good estimates on the coefficients
of powers of $s^{1/2}$ and $s\log s$ up to a given order $k$ one needs to consider $n\gg k$. Such
calculations, that can be performed with symbolic computer calculus, are beyond the scope of this
work and will be left to a future investigation. What is important to stress here is the crucial
difference between case (A) and case (B) upon iteration of the chain computation to fulfill
eq.~\eqref{eq:nth}. Indeed, eq.~\eqref{eq:hF2hnu} and the integration rule \eqref{eq:myint} imply
that: 
\begin{description}
\item{(A) }Given two integers $p$ and $q$, a term $s^{p/2}\log^q\!s$ in $h_F(s)$ produces in
  $h_\nu(u)$, besides a whole analytic function of $u$, a term $u^{p'}\log^{q'}\!u$ where $p'=p+1$
  always while $q'=q$ if $p$ is even and $q'=q+1$ if $p$ is odd; hence if $q\le \tfrac12p$, then
  $q'\le\tfrac12p'$. 
\item{(B) }In this case $p'=p$ always while $q'=q$ if $p$ is odd and $q'=q+1$ if $p$ is even.
\end{description}
It then follows that:
\begin{description}
\item{(A) }If the expansion of $h_F(s)$ contains only terms with powers and logs that satisfy
  $q\le\tfrac12p$, then so does the expansion of $h_\nu(u)$; in turn, $h_x(u)$ admits a similar
  expansion even upon all power and log multiplications implied by eq.~\eqref{eq:pois2} (this can be
  made more apparent by the change of variable $u=t^2/2$); finally the same conclusion applies to
  $h_G(s)$ since no new singularity can be introduced by the last step \eqref{eq:x2G} of the
  chain. Since at the lowest orders we have $(p,q)=(1,0)$ or $(p,q)=(2,1)$, then $q\le\tfrac12p$ at
  any order and there is no proliferation of logarithms. In other words, all four functions $h_F$,
  $h_\nu$, $h_x$ and $h_G$ admit a formal expansion in powers and logs of the form
  \begin{equation}\label{eq:asympt}
    h(s) = \sum_{n=0}^\infty\sum_{k=0}^n\big(a_{nk}+b_{nk}s^{1/2}\big)s^n\log^{n-k}\!s \;,
  \end{equation}
  where $a_{00}=0$ by construction. With the iterative procedure we can determine for instance that
  \begin{equation*}
    b_{F,00} = - \tfrac{32}{45} K \;,\quad a_{F,10} = -\tfrac2{45}K^2 \;,
  \end{equation*}
  but we cannot compute the other coefficients, since they are all coupled by the chain
  \eqref{eq:chain}.

\item{(B) }In this case nothing stops the logarithm proliferation which takes place already at the
  first order in $s$ with the appearance of terms $(p,q)=(2,n)$, $n=0,1,2\ldots$; this proliferation
  can only worsen at higher orders.
\end{description}

The difficulty with logarithms of case (B) could be a shortcoming only of the iterative asymptotic
analysis, rather than an indication that the original problem in eq.~\eqref{eq:FG} or
eq.~\eqref{eq:hhh2} does not really admit the $1/r-$cusped solution of case (B). Beyond the
iteration scheme, that is assuming that resummation of all logarithms does take place at any order
$s^n$, we could envisage four cases in increasing order of complexity/singularity of $h_F(s)$ as
$s\to0$:
\begin{description} 
\item{(B1)} $h_F(s)$ tends to $0$ as an infinitesimal of order less than $1$ but with a definite sign;
\item{(B2)} $h_F(s)$ is an infinitesimal of order less than $1$ but oscillates indefinitely around
  $0$;
\item{(B3)} $h_F(s)$ does not tend to $0$ and oscillates indefinitely around $0$, but remains
  bounded so that $1+h_F(s)$ stays positive; this, strictly speaking, would violate the leading
  asymptotics in eq.~\eqref{eq:lead} or eq.~\eqref{eq:hhh3};
\item{(B4)} $h_F(s)$ is unbounded as $s\to0$.
\end{description}
Case (B4) actually means that there no case (B) to start with: if (B4) holds true, the DARKexp model
admits only the cored density profiles of case (A). Case (B2) and (B3) are very difficult to
ascertain analytically; perhaps the oscillations in the slope of the density profile reported in
ref.~\cite{hjowill2}, on the basis of an essentially undisclosed numerical method, are a
manifestation of this fact. Here we rule out the smoothest case (B1), by showing it to be
incompatible with eq.~\eqref{eq:nth}.

Assuming that $h_F(s)$ has property (B1) means that there exists a number $\alpha$, with
$0\le\alpha<1$, such that $\lim_{s\to0}h_F(s)/s^{\alpha_-}=0$, $\forall \alpha_-<\alpha$ whilst
$\lim_{s\to0}|h_F(s)|/s^{\alpha_+}=+\infty$, $\forall \alpha_+>\alpha$. This can be reformulated in
a more convenient way as
\begin{equation}\label{eq:moreconv}
  h_F(st) = h_F(s)t^\alpha + o(h_F(s))\;, \quad \forall t>0 \;, \quad {\rm as}\;s\to0 \;.
\end{equation}
Then eq.~\eqref{eq:nth} implies
\begin{equation}\label{eq:hGhF}
     h_F(s) + h_G(s) \simeq 0\;,
\end{equation}
since the quadratic term $h_F(s)h_G(s)$ is of higher order and can be neglected. Let us now restart
the functional chain \eqref{eq:chain}. By the (B) case of eq.~\eqref{eq:hF2hnu} and by
eq.~\eqref{eq:moreconv} we have
\begin{equation*}
  \begin{split}
    h_\nu(u) &= \tfrac32 u\!\int_u^1\!ds\,[1+h_F(s)]s^{-5/2}\,(s-u)^{1/2} -1 \\
    & = (1-u)^{3/2} - 1 + \tfrac32 \int_0^{1-u}\!dt\,t^{1/2}\,h_F(u[1-t]^{-1})
    \simeq \tfrac32 B(\tfrac32,1-\alpha)\,h_F(u) \;,
  \end{split}
\end{equation*}
Then Poisson's eq.~\eqref{eq:pois2} yields
\begin{equation*}
  h_x(u) \simeq - \frac{3 B(\tfrac32,1-\alpha)}{\alpha(\alpha+3)+4}  h_F(u) \;, 
\end{equation*}
where we used the derivatives w.r.t. $t$ of eq.~\eqref{eq:moreconv} to estimate $h_F'(u)$ and
$h_F''(u)$. Finally, by eqs.~\eqref{eq:x2G} and \eqref{eq:moreconv}:
\begin{equation*} 
  \begin{split}
    h_G(s) &= \tfrac{35}{32}s^{-7/2}\int_0^s\!du\,u^3[1+h_x(s)]^3(s-u)^{-1/2} -1 \\
    &\simeq \tfrac{105}{32}\int_0^1\!dt\,h_x(st)\,t^3(1-t)^{-1/2} \simeq 
    \tfrac{105}{32} B(\tfrac12,\alpha+4)h_x(s) \;. 
  \end{split}
\end{equation*}
Hence
\begin{equation*}
   h_F(s) + h_G(s) \simeq h_F(s) \left[ 1 - \frac{305}{32} 
   \frac{B(\tfrac12,\alpha+4) B(\tfrac32,1-\alpha)}{\alpha(\alpha+3)+4} \right]
\end{equation*} 
which can be verified to falsify eq.~\eqref{eq:hGhF}, since the quantity between square
brackets in the r.h.s. is strictly negative for $0\le\alpha<1$.

\section{The cored profiles near the origin}\label{sec:corig}

As shown in the previous section, the situation for case (A) of cored density profiles is
much more under control. We have established that, as $s,u\to0$,
\begin{equation}\label{eq:asycore}
  \begin{split}
    F(s) &= K s^{-1}\left[1 - \tfrac{32}{45} K s^{1/2} -\tfrac2{45}K^2s\log s
      +O(s)\right] \;,\\[1ex]
    \nu(u) &= 3\left[1 - \tfrac13\pi K u^{1/2} - \tfrac{16}{135}K^2u\log u 
      +O(u) )\right] \;,\\[1ex] 
    x(u) &= (2u)^{1/2}\left[1 + \tfrac{1}{12}\pi K u^{1/2} + \tfrac4{225}K^2u\log u 
      +O(u)\right] \;,\\[1ex]
    G(s) &= \tfrac{\pi}{4\sqrt2}s^2\left[1 + \tfrac{32}{45} K s^{1/2} +\tfrac2{45}K^2
      s\log s +O(s)\right] \;, 
  \end{split}
\end{equation}
which are the beginning of full expansions of the form \eqref{eq:asympt}. The constant $K$ cannot
be determined by the asymptotics analysis, since it depends on the full structure of the
solution. The main equation \eqref{eq:FG} of the DARKexp model fixes it in terms of the unique, {\it
  temperature-like} parameter $b$ of the model and the dimensionless mass of the system:
\begin{equation*}
  K = \frac{2^{5/2}\, b^2m}{\pi(e^b-1-b)}
\end{equation*}
By the (A) case of eq.~\eqref{eq:Km}, the full solution of eq.~\eqref{eq:hhh2} would then
determine $K$ (and therefore also $m$) as a function of $b$.

\smallskip
Inverting the third of eqs.~\eqref{eq:asycore} yields
\begin{equation*}
  u(x) = \tfrac12x^2\left[1 - \tfrac{1}{6\sqrt2} \pi K x - \tfrac8{225}K^2x^2\log x 
      +O(x^2)\right] \;,
\end{equation*}
which inserted into the second of  eqs.~\eqref{eq:asycore} entails
\begin{equation*}
  \nu(u(x)) = 3 - \tfrac{1}{\sqrt2} \pi K x - \tfrac{16}{45}K^2x^2\log x 
  +O(x^2) \;.
\end{equation*}
Let us also recall that the dimension--full density and potential are recoved as 
\begin{equation*}
   \rho(r) = \tfrac13\rho_0\nu(u(r/r_\ast))   \;,\quad
   \phi(r) = \tfrac43\pi G\rho_0r^2_\ast\big[-1 + u(r/r_\ast)\big]  \;.
\end{equation*}
Therefore we have
\begin{equation*}
  \gamma_0 \equiv \frac{\rho'(0)}{\rho(0)} = - \frac{\pi K}{3\sqrt2 \,r_\ast} 
\end{equation*}
and we can absorb our ignorance of $K$ in $\gamma_0$, which can be regarded as the second, and last,
free scale parameter in place of $r_\ast$ (the first free scale parameter is $\rho_0$ itself). Hence
we can substitute $K$ also in the logarithmic term and obtain
\begin{equation}\label{eq:rhonear}
  \rho(r) \simeq \rho_0\left[1 + \gamma_0r - 
    \frac{\!\!32}{5\pi^2}\,\gamma_0^2\,r^2\log r + O(r^2)\right]\;.
\end{equation}
Quite interestingly, just because of the linear term, when viewed in three dimensions the density
exhibits a conical shape in the neighborhood of the origin as the Burkert profile.

\section{Conclusions and outlook}

The main result of the present investigation is that the DARKexp model allows for cored density
profiles in fully relaxed halos of purely self--gravitating matter. We have established the
analytic structure of the cored profiles in the asymptotic sense, that is without any control on the
convergence properties of expansions like that in eq.~\eqref{eq:asympt}. We have also determined the
first two coefficients of those expansions in terms of the global normalization constant $K$ defined
in eq.~\eqref{eq:Km}. Notice that this allows to extract explicit numbers in the ratios
$b_{00}^2/a_{10}$. For instance in the case of the density profile we have
\begin{equation*}
  \frac{b_{\nu,00}^2}{a_{\nu,10}} = -\frac{15}{16}\pi^2 \;,
\end{equation*}
a fact already used to obtain eq.~\eqref{eq:rhonear} in the previous section.
 
Our control of the DARKexp problem, that is the determination of the dimensionless phase--space
distribution $F(s)$ from eq.~\eqref{eq:FG}, is much more limited when $1/r-$cusps are assumed, as
done in ref. \cite{hjowill1} where the DARKexp was first put forward. We have uncovered an unbounded
proliferation of logarithms in the iterative asymptotic analysis when $F(s)\sim s^{-5/2}$ to leading
order as $s\to0$, but we could establish very little about any possible resummation of such
logarithms. We only excluded the simplest possibility (B1) of section \ref{sec:sublead}, namely we
ruled out that $F(s)\sim s^{-5/2}[1+O(s^\alpha)]$ for $0<\alpha\le1$. This means that $F(s)$ should
have an oscillatory behavior in $\log s$ as $s\to0$, perhaps not even a damped one. As consequence
the density $\rho(r)$ would exhibit oscillations in $\log r$ for small enough $r$. Something of this
kind has been reported in the literature about a numerical solution of the DARKexp problem
\cite{hjowill2}. Density slope oscillations, that is oscillations in the logarithmic derivative of
$\rho(r)$ have been discussed from a more general point of view in ref.~\cite{younga}.

\medskip
Coming back to the principal subject of this work, that are cored density profiles in the DARKexp
model, one might wonder about the possible applications of our results to realistic cosmological or
astrophysical contexts. However, for the moment there is no true margin for applicability.

First of all, regardless of cusps or cores, the DARKexp model, as an implementation of the principle
of maximum entropy into the spherically symmetric collapse of continuous self--gravitating matter,
cannot say anything about the statistics of relaxed halos, as realized in the Universe or in $N-$body
simulations. This statistics certainly retains information of the initial conditions as well as the
history of structure formation (and maybe also on the mass of the DM particle). Secondly, as any
model based on an ergodic phase--space distribution, the DARKexp model has two free scale
parameters, which we chose to be $\rho_0$ and $r_\ast$ (or $\gamma_0$), so that cores could be made
as small as one wants by hand. Hence the simplest way to check the model against $N-$body
simulations is to directly compare the differential energy distribution $n(E)$, as positively done
in refs.~\cite{hjowill3,nolting16}. But in this way no preference between cusps or cores may ever
arise. Such a preference could arise from a comparison of the mass density itself, in observations
or $N-$body simulations, as done in refs. \cite{beraldo13,umetzu16,nolting16}. The problem is in the
low resolution, that might confuse small cores with cusps. Indeed, any comparison of this type
requires to fix a distance scale from directly measurable quantities, such as, for instance
$r_{-2}$, the largest radius at which the logarithmic slope $r\rho'(r)/\rho(r)$ takes the value
$-2$; but this requires to go far beyond the asymptotic behavior studied in this work.

In conclusion, effective comparisons to observations and $N-$body simulations require the full
DARKexp cored density profile. In turn, this calls for a numerical approach capable to smoothly
connect to the cored asymptotics. Furthermore, the role of angular momentum should be clarified for
cored solutions, as done for $1/r$ cusps in ref. \cite{hjowill4}. These are two directions for
further improvements on cored systems in the DARKexp model.

\acknowledgments
The author aknowledges the contribution of Michele Turelli in the early stages of this work.


\end{document}